\documentclass[a4paper,11pt]{article}
\usepackage{pos}

\title{Narrow Spectra in  Heavy Quark Systems}

\author*[a]{Estia Eichten}
\affiliation[a]{Fermi National Accelerator Laboratory, \\
Batavia, Illinois, 60510 USA}

\emailAdd{eichten@fnal.gov: ORCID 0000-0003-0532-2300 \\ FERMILAB-CONF-24-1013-T (Approved)}

\abstract{Opportunities to observe new hadrons with very narrow widths are discussed.  Two areas of focus are the lowest P-states of heavy-light mesons and doubly heavy baryons.}

\FullConference{The XVIth Quark Confinement and the Hadron Spectrum Conference (QCHSC24)\\
 19-24 August, 2024\\
 Cairns Convention Centre, Cairns, Queensland, Australia\\}

\begin{document}
\maketitle

\section{Overview}

In the recent years many heavy quark states with very narrow widths ($\leq $ MeV) have been observed.  The properties of such states have given detailed insight into the
dynamics of QCD.  Such narrow states can't be far about the threshold for Zweig allowed strong decays.  We can classify the opportunities for such narrow states  by 
their heavy quark content as follows:
 \begin{itemize}
\item{$c\bar{c}, b\bar{b}, b\bar{c}, ccc, ccb, cbb {~\rm and~} bbb$ systems. Very narrow states are possible below  the respective thresholds the Zweig allowed decays (e.g, $D\bar{D}$ in the $c\bar{c}$ system, etc). Some resonance states very near threshold might be narrow. Many such states have already been observed \cite{ParticleDataGroup:2024cfk} and the predictions for the rest are well established.}
\item{Heavy-light mesons ($\bar{c}q, \bar{b}q$) q = u,d,s. Generally strong decays above
ground states and possibly some of the lowest P-states.}
\item {Doubly heavy baryons : $ccq, bbq,\{bc\}q {~\rm and~} [bc]q$. For unequal mass heavy quarks, two set of states exist depending on whether the heavy quarks are in a symmetric (\{\}) 
or an antisymmetric ([]) spatial configuration. Can any excited states be narrow?}
\item{Tetraquark systems with heavy quarks (c or b) plus light degrees of freedom. Here, surprisingly, some ground states are actually stable against strong decays\cite{Eichten:2017ffp}.}
\item{Tetraquark systems with all heavy quarks. Near threshold wide  $c\bar{c}c\bar{c}$ resonance  states have been observed at the LHC\cite{CMS:2023owd}. There might be related  narrow tetraquark states.}
\end{itemize}
 How many more such states remain to be found?  
In the remaining sections the prospects for discovery are examined.

\section{Heavy Light Mesons}

Consider a system consisting of both heavy and light quarks.  The total angular momentum of the light subsystem $\bf{j}_l$ is just the sum of the total orbital angular momentum, $\bf{l}$,
and the total spin, $\bf{s}$ of that subsystem,  ie ($\bf{j}_l=\bf{l}+\bf{s}$). Similarly, the total angular momentum of the heavy subsystem is just $\bf{ J}_h=\bf{L}_h+\bf{S}_h$.
 Thus the total angular momentum is $\bf{J}= \bf{j}_l+\bf{J}_h$.  
 In the case of  heavy-light mesons considered in this section, the heavy quark is nearly static, so $\bf{J}_h = \bf{S}_h$.
 
 The two systems of heavy-light mesons observed to date are the  $D_{u,d}, D_s$ and $B_{u,d}, B_s$ mesons.  All the ground states ($L=0$) $J^P=0^-,1^-$ are stable against strong decays and are very narrow.  All the excited states above the lowest P states  have strong decays.  
 
 The lowest P states ($L=1$) have $\bf{j}_l = \frac{1}{2}$ or $\bf{\frac{3}{2}}$ and  positive parity. 
 The two $\bf{j}_1=\frac{1}{2}$ states are $\bf{J}^P= 0^+,1^+$ and the $\bf{j}_l=\frac{3}{2}$ states are $ \bf{J}^P= 1^+,2^+$. These P states are a special case because strong decay to the 
 associated two ground states $\bf{j}_1=\frac{1}{2}$ with $ \bf{J}^P= 0^-,1^-$ S waves might be kinematically forbiden. 
 \subsection{D mesons}
  The mass splittings in the $D_s$  system between the  $\bf{j}_l =\frac{1}{2}$ and $\bf{j}_l =\frac{3}{2}$ P states was very surprising when the $\bf{j}_l =\frac{1}{2}$ states were first observed
 experimentally.  These  $\bf{j}_l =\frac{1}{2}$ masses violated the expectations of simple potential models.  Various theoretical approaches were proposed to understand this 
 unexpected behaviour. These approaches generally followed the directions:  (1) Treat the ground  $\bf{j}_l =\frac{1}{2} (P=-1)$  S states and $\bf{j}_l =\frac{1}{2} (P=+1)$  P states as a chiral multiplet (e.g. \cite{Bardeen:2003kt}),  (2)  Treat the $\bf{j}_l =\frac{1}{2} (P=+1)$  P states  as bound states in $KD$ and $KD^*$  scattering\ (e.g. \cite{Guo:2006fu, Guo:2006rp,Guo:2009ct}), or (3) Consider  these states  as a mixture with both a quarkonium core and molecular component. 
 For a recent effort in this dirrection see e.g.  \cite{Zhang:2024usz}). %

\begin{figure}[htbp]
\begin{center}
\includegraphics[width=4in]{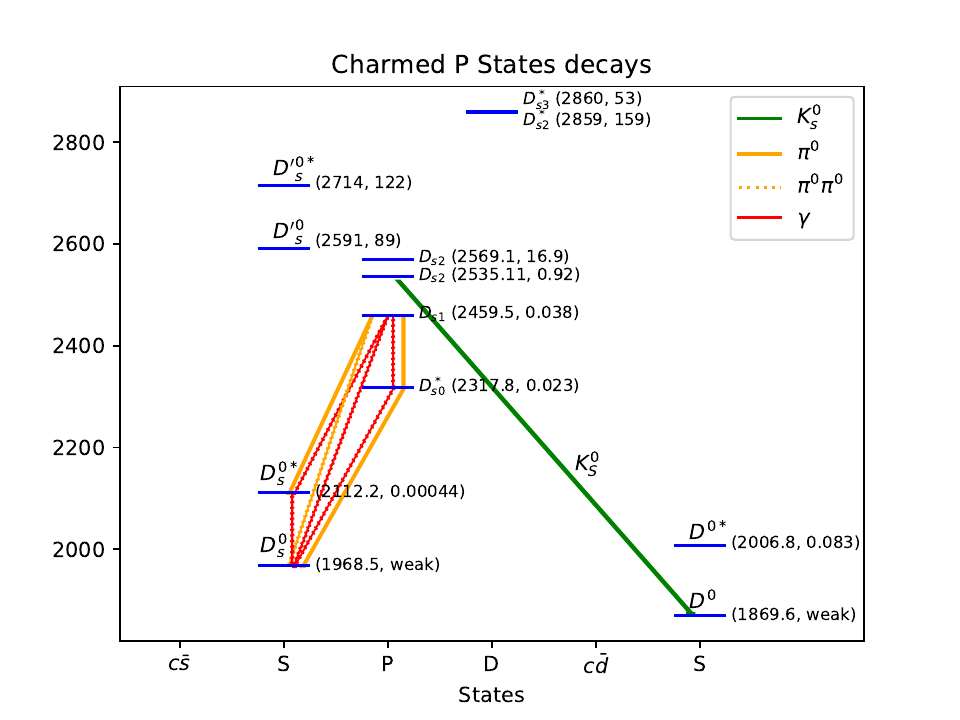}
\caption{The predicted hadronic and electromagnetic transition rates for
narrow  $1S$  $ j_l^P = 1/2^-$ and $1P $ $ j_l^P = 1/2^+$ in ($c\bar{s}$) heavy-light states.} 
\label{ctran}
\end{center}
\end{figure}

For $D_{u,d}$ states,  the $\bf{j}_l =\frac{1}{2}$ P states have  total widths \cite{ParticleDataGroup:2024cfk} $\Gamma[{D^*_0(2343)}]=  229\pm 16$ MeV and $\Gamma[{D_1(2412)})]=  314\pm 29$ MeV 
 and  have S wave  pion transitions  and  $\Gamma[{D_1(2422)}] =  31.3\pm 1.9$ MeV and  $\Gamma[{D_1(2412)}] =  314\pm 29$ MeV and 
$\Gamma[{D^*_2(2461)}] =  47.3\pm 0.8$ MeV for the $\bf{j}_l =\frac{3}{2}$ states which have  D wave transitions.
 For $D_s$  state,  $\Gamma[{D_{s1}(2535}] =  0.92\pm 0.05$ MeV and $\Gamma[{D^*_{s2}(2569)}] =  16.0 \pm 0.7$ MeV for the $\bf{j}_l =\frac{3}{2}$ states 
 which have  D wave  kaon transitions.  
 The $\bf{j}_l = \frac{1}{2}$ states are below the kinematic threshold for Zweig allowed strong decays. Their transitions  are shown in Figure \ref{ctran}
 Detailed calculations of the various transition rates for these states are model dependent. Measurements of the rates can distinguish 
 between  theoretical approaches. Table \ref{tbl:width-exp} shows some examples.
\begin{table*}
\begin{center}
\caption{The predicted hadronic and electromagnetic transition rates for
narrow  $1S j_l^P = 1/2^-$ and $1P j_l^P = 1/2^+$ heavy-light states. The original model of Bardeen et al.\cite{Bardeen:2003kt} $c\bar{s}$ 
model along with a more recent relativistic model \cite{Han:2023wqq}and a model including coupling to decay channels \cite{Aaij:2017ueg} (using $\Lambda' =1.0$ (GeV)) are compared with 2024 PDG decay rates for the $D_s$ system \cite{ParticleDataGroup:2024cfk}.
\label{tbl:width-exp}}
\begin{tabular}{clcccl} 
\hline
\hline
system &  transition & & partial widths (kev) & &   exp BR   \\
&   &  \cite{Bardeen:2003kt}(2003) & \cite{Han:2023wqq}(2023) & \cite{Aaij:2017ueg}(2024)  & PDG \cite{ParticleDataGroup:2024cfk} (2024) \\
\hline
\hline
$(c\bar u)$ &  $1^- \rightarrow 0^- + \gamma$  & $33.5$ &    &      & ($38.1\pm 2.9$)\% \\
            &  $1^- \rightarrow 0^- + \pi^0$             & $43.6$  &    &      & ($61.9\pm 2.9$)\% \\
            &  ~~~~~total                                         &   $76.9$ &    &       &   \\
\hline
$(c\bar d)$ &  $1^- \rightarrow 0^- + \gamma$  &  $1.63$ &    &     & ($1.6\pm 0.4$)\%     \\
            &  $1^- \rightarrow 0^- + \pi^0$             & $30.1$   &    &    &  ($30.7\pm 0.5$)\%  \\
            &  $1^- \rightarrow 0^- + \pi^+$            &  $65.1$   &    &    & ($67.7\pm 0.5$)\%   \\
            &  ~~~~~total                                         &  $95.8$  &     &       &  $96\pm 22$             \\
\hline
$(c\bar s)$ &  $1^- \rightarrow 0^- + \gamma$  & $0.43$ &      &    &  ($93.6\pm 0.4$)\%    \\
            &  $1^- \rightarrow 0^- + \pi^0$  &       $0.0079$  &      &    &  ($5.77\pm0.35$)\%     \\
            &  ~~~~~total                     &                     $0.44$ &       &    &                                 \\
\hline
$(c\bar s)$ &  $0^+ \rightarrow 1^- + \gamma$ & $1.74$ &  ($2.55\pm ^{+0.37}_{-0.45}$  & 1.41  &   $(<5)$\%              \\
            &  $0^+ \rightarrow 0^- + \pi^0$           & $21.5$ &  ($7.83^{+1.97}_{-1.55}$)    &   & $(100~^{+0 }_{20})$\%   \\
            &  ~~~~~total                                        & $23.2$ &       &     &                               \\
\hline
$(c\bar s)$ &  $1^+ \rightarrow 0^+ + \gamma$  & $0.43$ &     &          & $(3.7^{+5.0}_{-2.4})$\%      \\
            &  $1^+ \rightarrow 1^- + \gamma$        &  $3.49$ &     & 2.19   &  ($< 8$)\%             \\  
            &  $1^+ \rightarrow 0^- + \gamma$        &  $7.62$ &     & 6.41    & ($18\pm 4$)\%          \\
            &  $1^+ \rightarrow 1^- + \pi^0$               & $21.5$ &   &     &    ($48\pm 11$)\%       \\
            &  $1^+ \rightarrow 1^- + 2\pi$                  & $ 9.7$  &   &      &  ($4.3\pm 1.3$)\%      \\
            &  ~~~~~total                                            & $42.7$ &    &      &                           \\
\hline
$(b\bar u)$ &  $1^- \rightarrow 0^- + \gamma$ & $0.78$  &     &      &    (seen)              \\
            &  ~~~~~total                                          & $0.78$ &      &      &                           \\        
\hline
$(b\bar d)$ &  $1^- \rightarrow 0^- + \gamma$ &  $0.24$ &      &      &   (seen)              \\   
            &  ~~~~~total                                         & $0.24$  &      &      &                           \\   
\hline
$(b\bar s)$ &  $1^- \rightarrow 0^- + \gamma$ & $0.15$ &       &      &                           \\   
            &  ~~~~~total                                         & $0.15$ &       &      &                           \\   
\hline
$(b\bar s)$ &  $0^+ \rightarrow 1^- + \gamma$ & $58.3$ &      & 17.56    &                           \\   
            &  $0^+ \rightarrow 0^- + \pi^0$            & $21.5$ &      &      &                           \\   
            &  ~~~~~total                                         & $79.8$ &       &      &                           \\   
\hline
$(b\bar s)$ &  $1^+ \rightarrow 0^+ + \gamma$  & $0.15$ &     &      &                           \\   
            &  $1^+ \rightarrow 1^- + \gamma$         &  $42.3$ &     &  10.62     &                           \\   
            &  $1^+ \rightarrow 0^- + \gamma$         & $58.3$ &      &  17.27   &                           \\   
            &  $1^+ \rightarrow 1^- + \pi^0$             & $21.5$ &      &      &                           \\   
            &  $1^+ \rightarrow 1^- + 2\pi$               & $0.24$ &      &      &                           \\   
            &  ~~~~~total                                         & $123.8$ &      &      &                           \\   
\hline
\hline
\end{tabular}
\end{center}
\end{table*}

\newpage
\subsection{B Mesons}
The 1P  $j^P =\frac{1}{2}^+$ states have not yet been observed.  If we knew the splitting in the limit of  the infinitely heavy quark,  then we could  use HQET to interpolate between the splittings in the$D_s$ system to predict the splitting in the $B_s$ system.  From a theoretical point of view what can we say about mass difference between the 1P $(j^P = 3/2^+)$ and $(j^P =1/2^+)$ meson states in QCD?  If one considers a light quark moving in a funnel potential about a static source the splitting of the form $A \bf{s}\cdot \bf{L}$.  u
Using a nonrelativistic Schrodinger eqution we have ($A>0$) while using a Dirac equation \cite{DiPierro:2001dwf} we have ( $A<0$).  Hence even the sign of the splitting is not known apriori.

\begin{figure}[htbp]
\begin{center}
\includegraphics[width=3in]{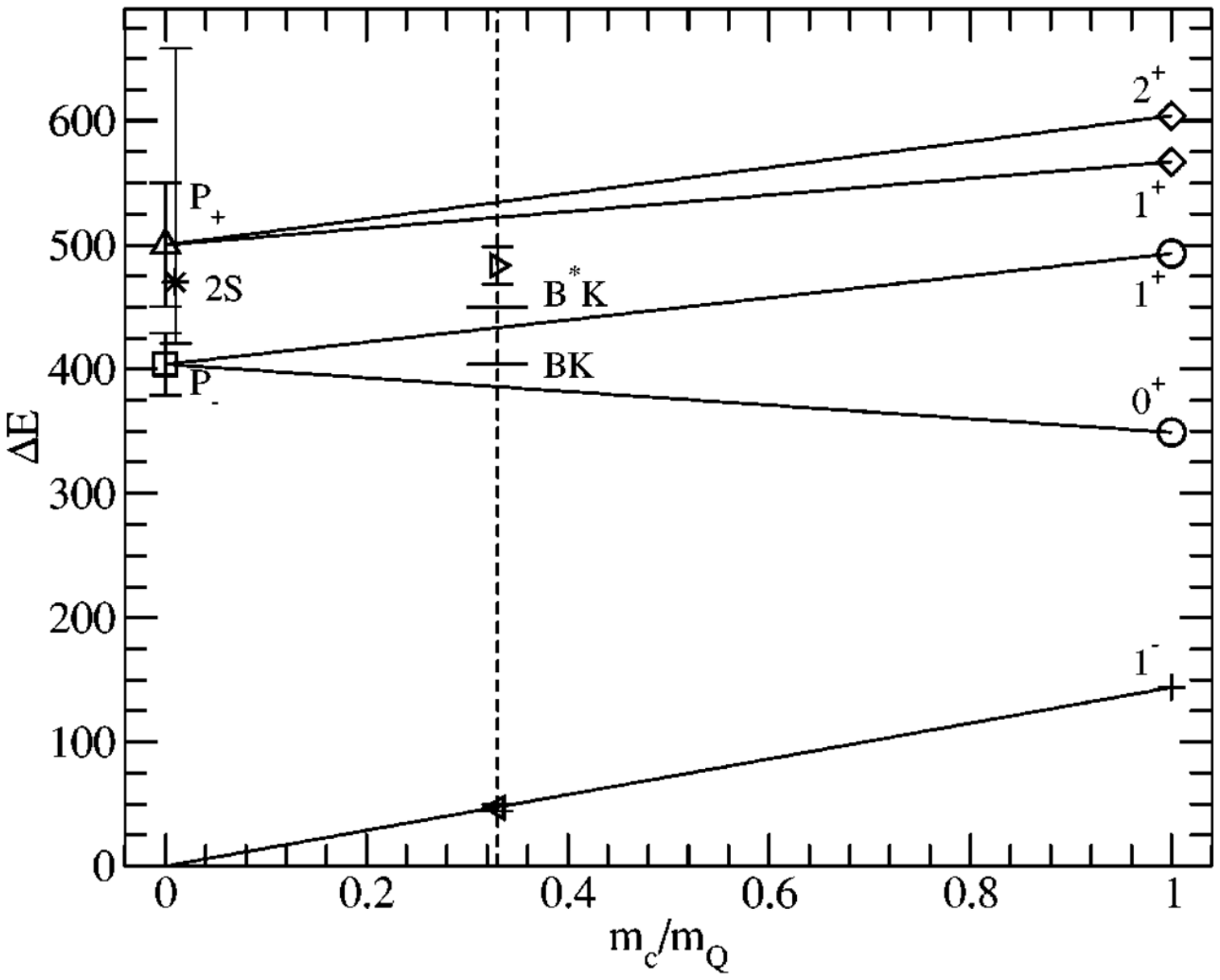}
\caption{The energies in MeV of P-wave excited states relative to
the ground state  heavy-light meson with a heavy quark
mass $m_Q$ and a light quark which is strange. Data from experiment
are plotted for charm and for b quarks while our lattice results are
shown for static quarks. The 2S excitation ~from their larger volume
results is also shown. The dotted vertical line gives the interpolated
value appropriate for b quarks. The BK and B*K thresholds are also
shown. 
These are the lightest isospin-conserving decay modes allowed by strong interactions. From Green et al. \cite{Green:2004}}
\label{fig:hqlimit}
\end{center}
\end{figure}

Lattice QCD results in the heavy quark limit have  been presented by Green et el.\cite{Green:2004}. The results for the 
interpolation between $\frac{1}{m_Q}=0$ and the charm-strange system is shown in Figure \ref{fig:hqlimit}.  Even more precise lattice.
 results  in the HQET limit could be useful.
Hence assuming linear behaviour in $1/m_Q$ and knowing M(1/2+) - M(1/2-) for ${1/m_Q \rightarrow 0}$ we can
predict the value $M(B_{s1})$ and $M(B^*_{s0})$.
However, including the effects of coupling to the strong decay channels or in
molecular model the distance of the state from the two body strong decay
threshold is critical. So one might not expect this simple behavior.

\begin{figure}[htbp]
\begin{center}
\includegraphics[width=4in]{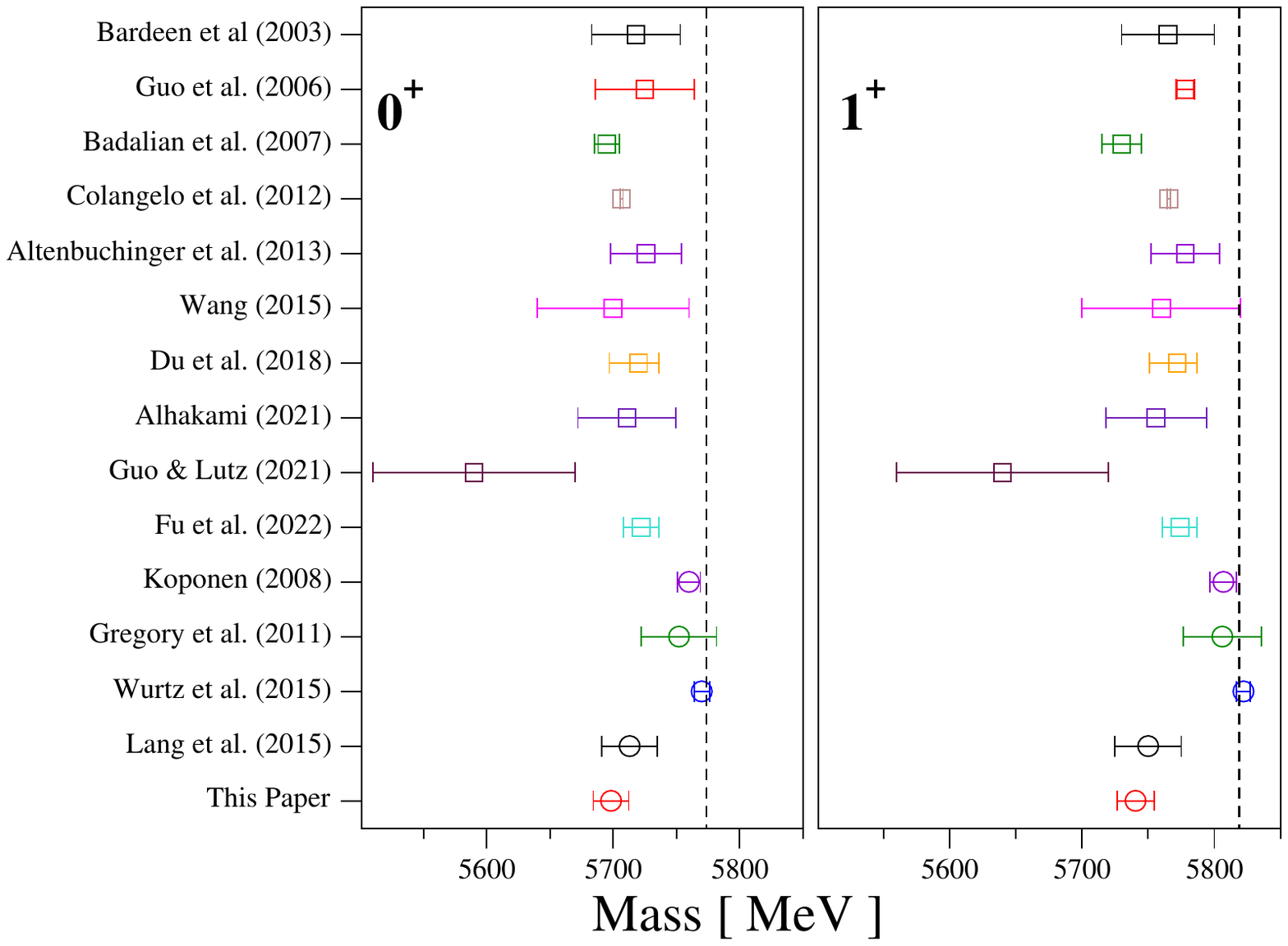}
\caption{Below Threshold}{Comparison of Hudspith and Mohler results \cite{Hudspith:2023loy} or the $B_{s0}^*$
(left pane) and $B_{s1}$ (right pane) ground-state masses
to various other results. Circles denote the results of lattice calculations, while squares denote the
results from model/EFT calculations. The vertical line denotes the respective threshold. To
translate their result for the binding energy to the result displayed in this plot an iso-symmetric
kaon mass of 494.2 MeV has been used. Adapted from\cite{Hudspith:2023loy}.}
\label{Hudspith}
\end{center}
\end{figure}

\begin{figure}[htbp]
\begin{center}
\includegraphics[width=4in]{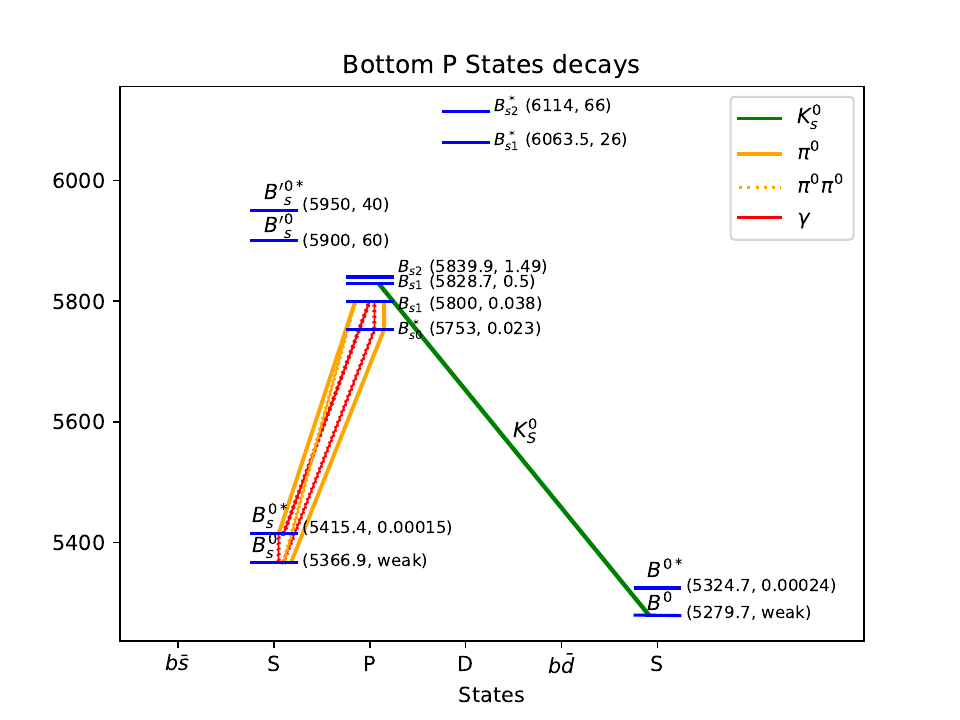}
\caption{The predicted hadronic and electromagnetic transition rates for
narrow  $1S$  $ j_l^P = 1/2^-$ and $1P $ $ j_l^P = 1/2^+$ in ($b\bar{s}$) heavy-light states.}
\label{btran}
\end{center}
\end{figure}

These masses can also be calculated  directly in Lattice QCD.  
Recently Hudspith and Mohler have presented a detailed LQCD calculation \cite{Hudspith:2023loy}. Their compare their results with earlier LQCD results and the results of other approaches.
The masses  obtained by various models are shown in Figure \ref{Hudspith}.  Essentially all the models show that the $j^+\frac{1}{2}$ states
are below the associated threshold for strong decays.  
If the masses of the $j^P=\frac{1}{2}^+$ P wave b mesons are as expected above, the allowed transitions would be given by Figure \ref{btran}.
\newpage
\section{Singly Heavy Baryons}
\begin{figure}[htbp]
    \centering
    \begin{minipage}{.45\textwidth}
        \centering
        \includegraphics[width=0.95\linewidth]{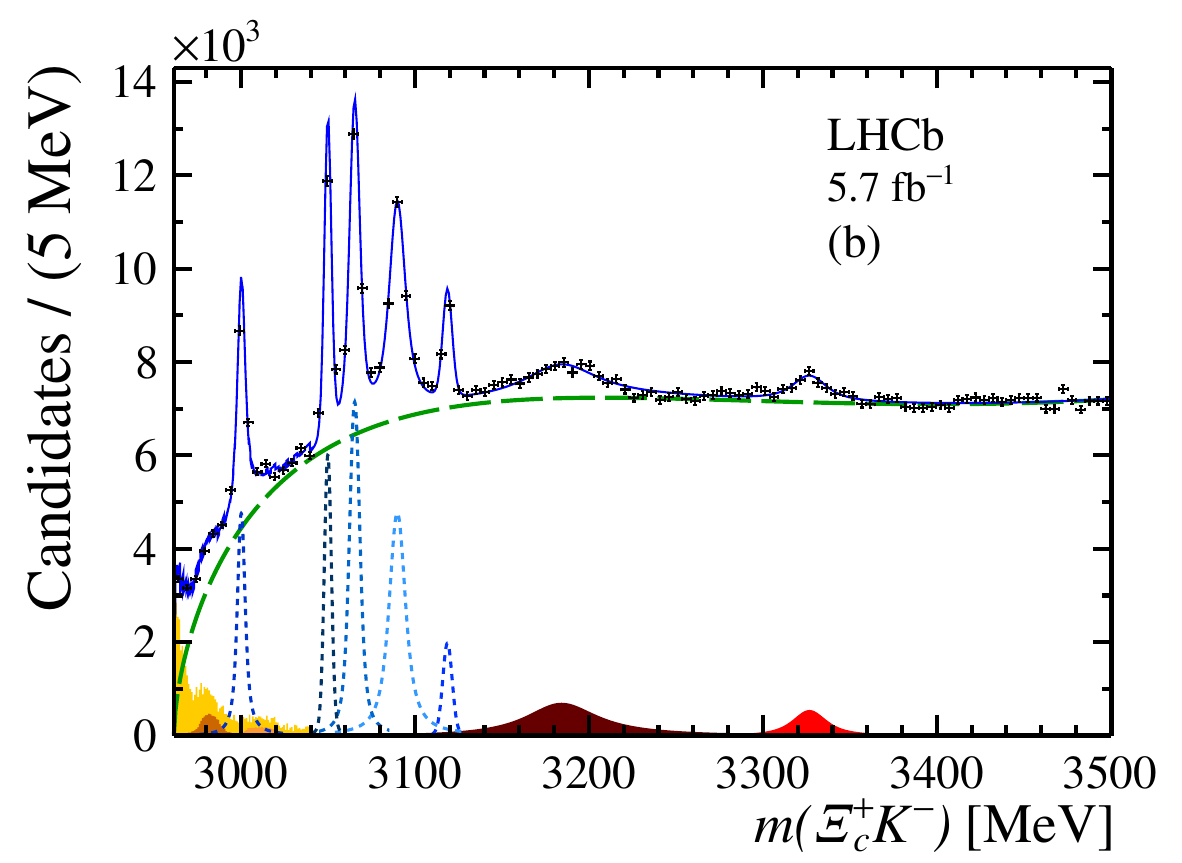}
    \end{minipage}%
    \begin{minipage}{0.55\textwidth}
        \centering
        \includegraphics[angle=270,width=0.95\linewidth]{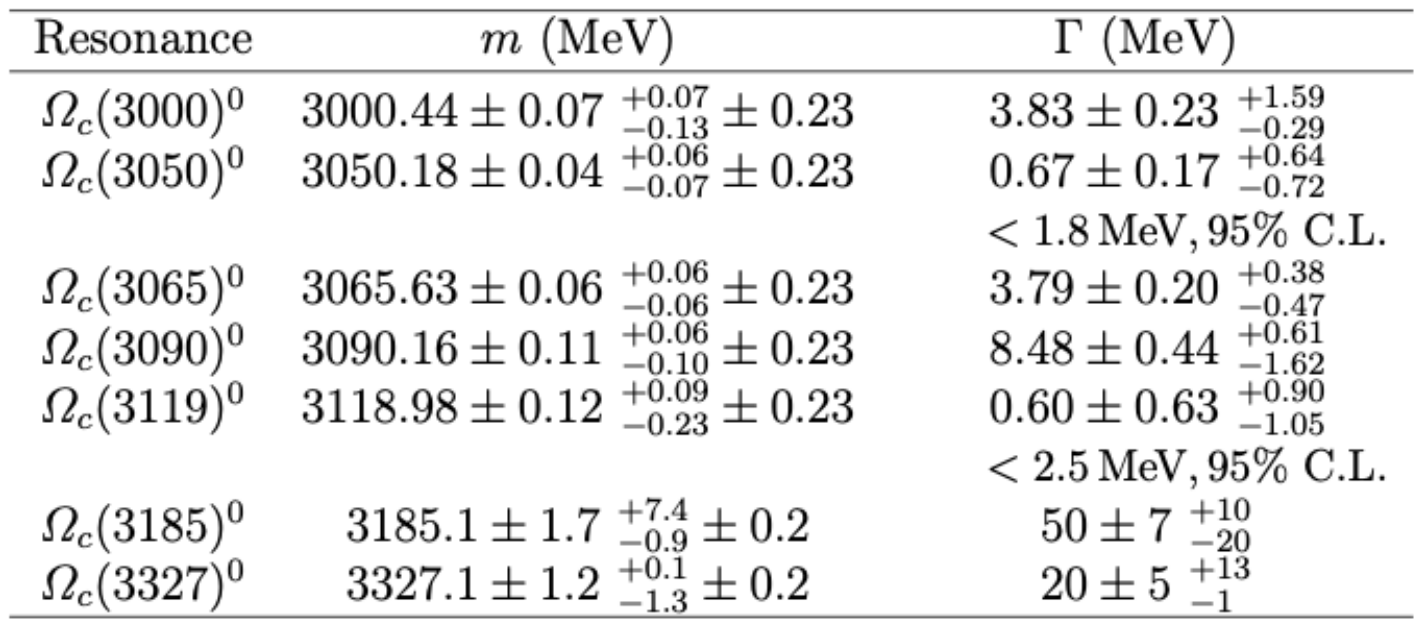}
    \end{minipage}
    \caption{The $\Xi_c^+ K^-$ invariant mass distributions  and measured masses and natural width of the  $\Omega_c(X)^0$  candidates \cite{LHCb:2024eyx}.    \label{fig:Omegac}}
\end{figure}

The lowest P states for heavy baryons have quantum number $J^P= \frac{1}{2}^-, \frac{3}{2}^-, \frac{1}{2}^-,\frac{3}{2}^-, \frac{5}{2}^-$.  
In fact the $\Omega_c$ lowest P wave states  Figure \ref{fig:Omegac} have been observed by LHCb \cite{LHCb:2017uwr} \cite{LHCb:2023sxp}. 
All the P state widths are fairly narrow which suggests that lowest $J^P$ state is not far about threshold for the strong decays 
$\Omega_c \rightarrow \Xi_c^+ K^-$. To date
 the quantum numbers have been determined for only for the $\Omega_c(3050)$ state \cite{LHCb:2024eyx}.
It is possible that for the still unobserved $\Omega_b$ analogies the $J^P$  P state might be below threshold for strong decays.
\section{Heavy Heavy Light Baryons}

For heavy-heavy-light baryons the only possibility for very narrow states above the ground states are the lowest P-wave states. Defining  $Q_1=c, Q_2= b$ and $q=u,d,s$ the baryons: $Q_iQ_jq$ 
the lowest P states for $q=u$ and $q=d$ have strong pion transitions to the associated  ground states.  while the  $Q_iQ_js$ P-waves masses may be low enough that K transitions to singly heavy 
baryons are kinematically forbidden.  These states would then be the analogy of the  lowest heavy-light meson P states.

\subsection{Models (in collaboration with  Chris Quigg)} 

One theoretically attractive picture of doubly heavy baryons is to consider the two heavy quarks form a subsystem similar to the NRQCD system that exists for $Q\bar Q$ quarkonium systems.
Heavy quarks $(Q_iQ_j)$ bound in a color $\mathbf{\bar 3}$ by an effective potential  of the ``Cornell'' Coulomb$\,+\,$linear form at half strength for both components. The strength of the Coulomb contribution is fixed by the color Casimir.  Lattice studies indicate that the effective string tension for the color $\mathbf{\bar 3}$ is half that for the singlet configuration \cite{Nakamura:2005hk}.  
Other lattices studies of the three body potential reach the same conclusion. 
For example, for baryons fitted with the $\Delta$ string configuration the string tension is $\sigma_{Q_i Q_j Q_k} \approx 0.53 ~ \sigma_{Q_i \bar{Q}_i }$ \cite{Takahashi:2002bw}.   
For sufficiently heavy $Q_i, Q_j$, it makes sense to regard the doubly heavy diquark and the nucleus of a diquark--light-quark atom, small in extent relative to overall size of the ``atom'' determined by the motion of the light (strange) quark. This system closely resembles a heavy--light meson with the same light degrees of freedom for the quantum states of the light quark~\cite{Savage:1990di}, with the important added element 
that the core can be excited.

In the heavy quark limit the average separation between the two heavy quarks (denoted $r$) is much smaller than the separation between the center of mass to the two heavy quark system and the light quark (denoted $R$) here.  In this limit the wavefunction for the baryon system the ground state can be factorized into the product of wavefunctions 
  \begin{equation}
   \Phi(r, R) = \phi(r)\psi(R) 
 \end{equation}
   where $\phi(r)$ is a solution of the NR Schroedinger  Equation for two color $3$ quarks combined into a total color $\bar 3$ .  
The solution of the system has excitation energies to is approximate one half the excitation energy of  the corresonding state in  $Q\bar Q$ as the solutions are not 
very dependent of $m_Q$ in charmonium to bottomonium region. 
For applications involving $b$ and $c$ heavy quarks, we will have to check that the planetary light quark orbiting a tiny diquark is a plausible approximation.
The low-lying excitation spectrum is given in Table \ref{tab:QQex}.
  \begin{table}[htbp]
    \centering
    \caption{The excitation energies (in MeV) of the low-lying excited states in the QQ systems compared with the $Q\bar{Q}$ systems.  Only states in a limited excitation energy  range (below 600MeV for the $bb$ system) are shown.}
     \vspace*{0.5cm}
    \begin{tabular}{|c|ccc|ccc|}
    \hline
     state & $cc$ & $bc$ & $bb$ & $\bar{c}c$ & $ \bar{b}c $ & $\bar{b}b$ \\
     \hline
      1P & 226 & 217 & 208 & 428 & 436 & 467 \\
      2S & 337& 311 & 278  & 591 & 570 & 563 \\
      1D & 393& 369 & 340 & 713 & 702 & 710 \\
      2P & 499 & 470 & 409 & 871 & 838 & 815\\
      1F & 537 &  498 & 448 & 951 & 919 & 898\\
      3S & 598  & 545 & 472 & 1015 & 957 & 902\\
      2D & 635 & 585 &  514 &1098 & 1046 & 980 \\
      1G & 666 & 615 & 544 & 1164 & 1110 & 1077 \\
      3P & 732 &  669 & 577 & 1242 & 1170 & 1095\\
      \hline
    \end{tabular}
    \label{tab:QQex}
 \end{table}
 
It is important to note that the splittings between the 1S and 1P state in the $[QQ]_{\bar{3}}$ subsystem are comparable to the splittings between the ground states and first excited states in 
a heavy-light meson.  Thus the spectrum of heavy-heavy-light baryons is not well approximated by two separate scales usually assumed for these baryons.  In particular we can compare the splittings in Table \ref{tab:QQex}.  
\subsection{A Simple Model}
\begin{figure}[htbp]
    \centering
    \begin{minipage}{.45\textwidth}
        \centering
        \includegraphics[width=0.95\textwidth]{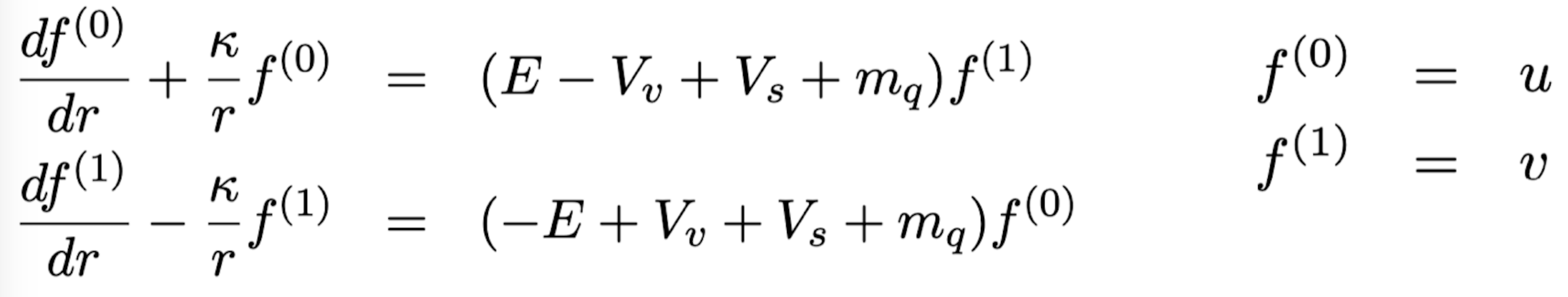}
    \end{minipage}
    \begin{minipage}{0.45\textwidth}
        \centering
        \includegraphics[width=0.95\textwidth]{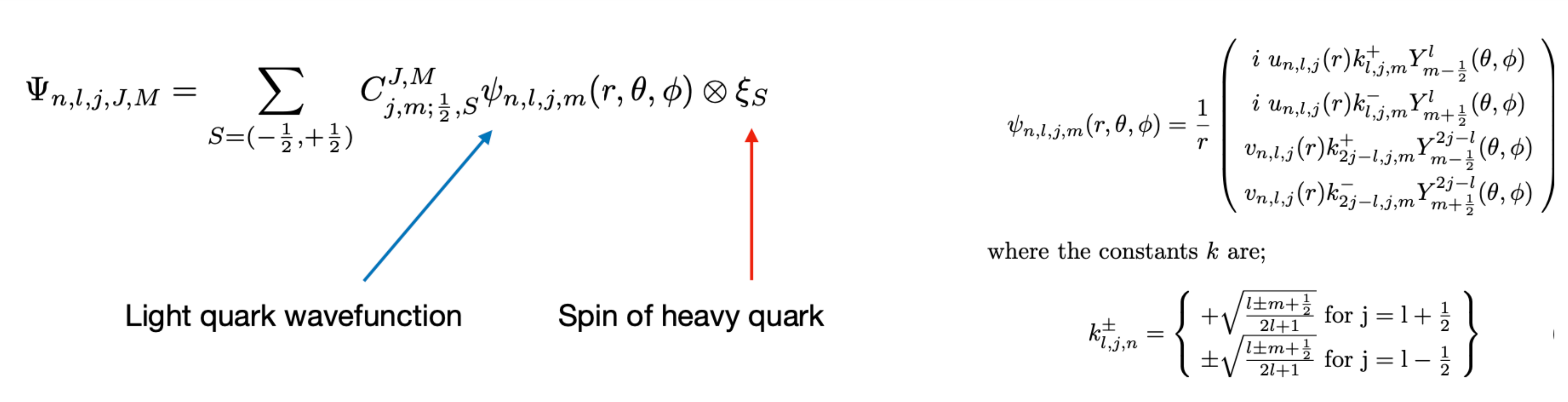}
    \end{minipage}
  \caption{Four component Dirac equations for the Cornell potential. Notation defined in \cite{DiPierro:2001dwf}.}
  \label{fig:dirac}
\end{figure}

One theoretically attractive picture of doubly heavy baryons is to consider the two heavy quarks form a subsystem similar to the NRQCD system that exists for $Q\bar Q$ quarkonium systems.
Heavy quarks $(Q_iQ_j)$ bound in a color $\mathbf{\bar 3}$ by an effective potential  of the ``Cornell'' Coulomb$\,+\,$linear form at half strength for both components. The strength of the Coulomb contribution is fixed by the color Casimir.  Lattice studies indicate that the effective string tension for the color $\mathbf{\bar 3}$ is half that for the singlet configuration \cite{Nakamura:2005hk}.  
Other lattices studies of the three body potential reach the same conclusion. 
For example, for baryons fitted with the $\Delta$ string configuration the string tension is $\sigma_{Q_i Q_j Q_k} \approx 0.53 ~ \sigma_{Q_i \bar{Q}_i }$ \cite{Takahashi:2002bw}.   
For sufficiently heavy $Q_i, Q_j$, it makes sense to regard the doubly heavy diquark and the nucleus of a diquark--light-quark atom, small in extent relative to overall size of the ``atom'' determined by the motion of the light (strange) quark. This system closely resembles a heavy--light meson with the same light degrees of freedom for the quantum states of the light quark~\cite{Savage:1990di}, with the important added element that the core can be excited. For applications involving $b$ and $c$ heavy quarks, we will have to check that the planetary light quark orbiting a tiny diquark is a plausible approximation. 
Such a two scale model is as follows:
\begin{itemize}
\item {$[(Q(r/2)Q(-r/2)]_{\bar{3}}$ :  NRQCD with $V(r) = - (2/3)\alpha /r+ r/(2 a^2) = 1/2 V(r) \rm{(Cornell)}$}
\item {$[QQ]_{\bar{3}} (0) s(R)$ :  Dirac equation for light quark motion around a static diquark.  Use a Cornell potential: $V_v =  -\frac{2}{3}\alpha/r$  and $V_s = \frac{r}{2a^2}$ the Dirac system of 
equations is shown in Figure \ref{fig:dirac}.}
\end{itemize}

\begin{figure}[htbp]
\begin{center}
\includegraphics[width=3in]{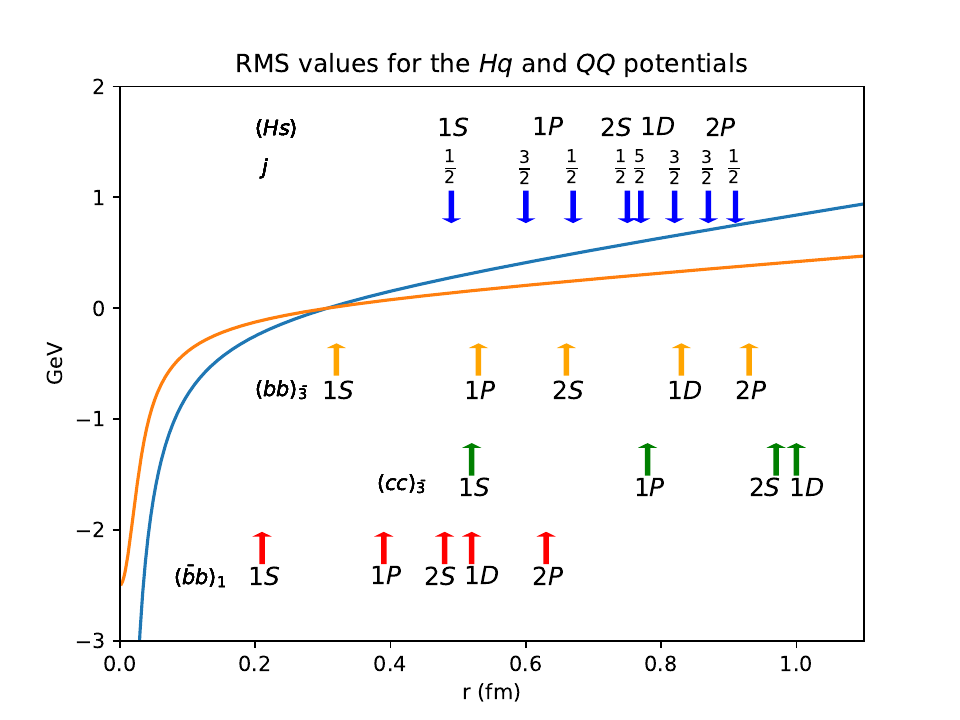}
\caption{Comparison of the RMS separation between the two heavy quarks  $QQ$ to the separation of  the heavy diquark system (H) and the strange quark  $q=s$ in various states for a doubly heavy baryon systems. For reference the RMS separation between between the $b$ and $\bar{b}$ in a bottonium systems is shown.}
\label{rms}
\end{center}
\end{figure}

Another way to see that the assumption of separation of scales fails for  heavy-heavy-light baryons (with $Q=c$ or $Q=b$) is to compute the RMS separations in the  the $Q_iQ_j$ system and the $QQ =H$ core  the light quark $q=s$ system .  This comparison using the Dirac equation for the $Hq$ system is shown in Figure \ref{rms}. 
\subsection{Born-Oppenheimer Effective Theory}
\begin{figure}[htbp]
\begin{center}
\includegraphics[width=2in]{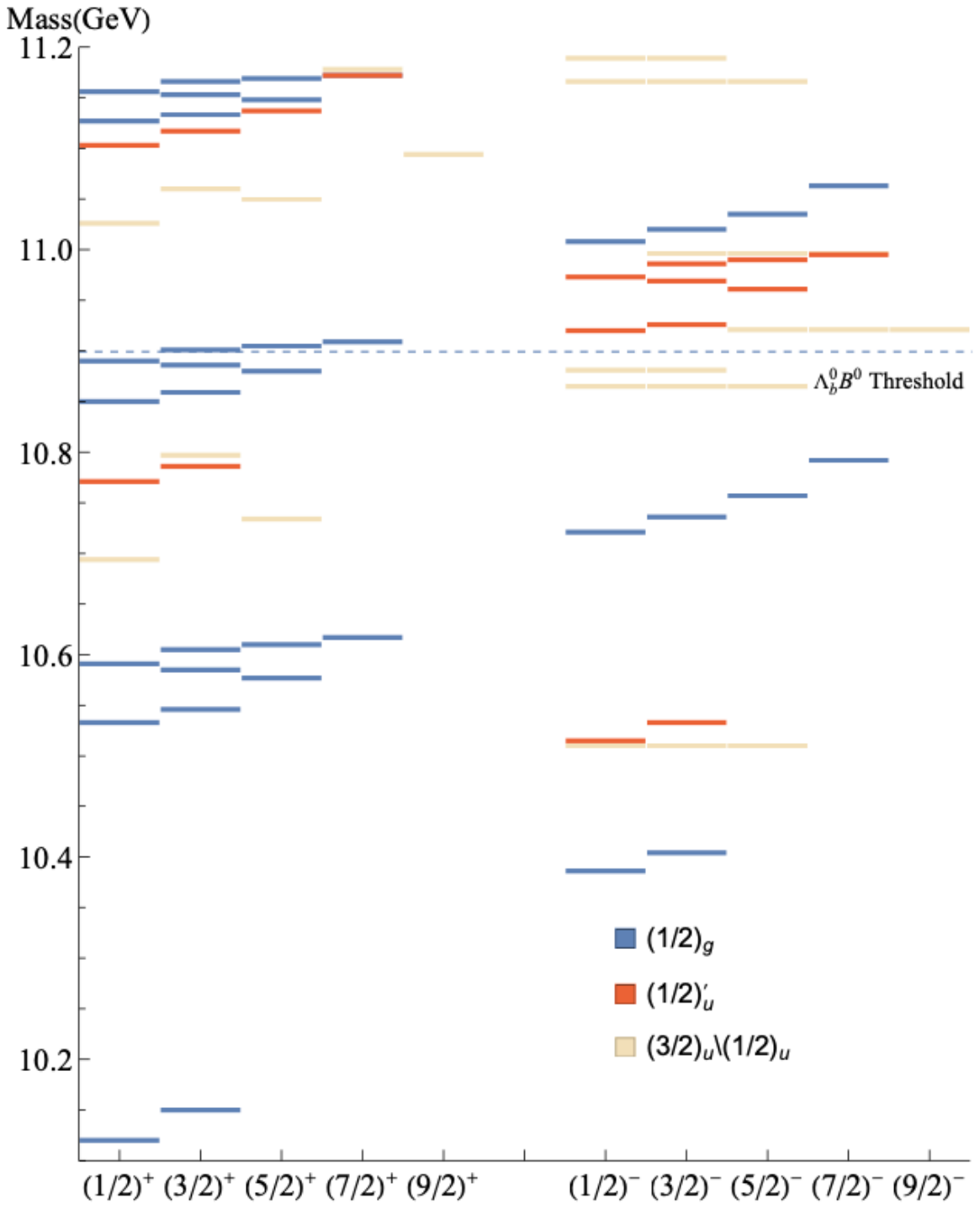}
\caption{Spectrum of double bottom baryons in terms of $j^{\eta_P}$ states. Each line represents a state. The spectrum is for the
states associated to the $(1/2)_g$ and $(1/2)_u'$ static energies and previous results  \cite{Soto:2020pfa} for
the mixed $(3/2)_u / (1/2)_u$ static energies, which do not include hyperfine contributions. The color indicates the static energies
that generate each state. From \cite{Soto:2021cgk}. \label{fig:Soto}}
\end{center}
\end{figure}

One can compute the lowest ($J=1/2,3/2,3/2,5/2$)   P wave states directly in LQCD.
Alternatively,  one can compute in Lattice QCD ground state energies E(R) for a system with one dynamic light quark in the static potential of two heavy static quarks separated by a distance R.
The potentials associated with $O(1/M_Q)$ corrections can also be calculated in this way.
Then the total energy and wavefuction for the heavy quark ground system can be obtained by solving the SE for each distinct set of quantum numbers. 
This allows the complete solution for the lowest doubly heavy  S and P states.  The Born-Oppenheimer approach for two heavy quarks can be calculated on the lattice \cite{Najjar:2009da}. Using this approach and a string potential \cite{Soto:2021cgk} one can computed the spectrum of doubly heavy baryons see Figure  \ref{fig:Soto}.

Both the simple potential model and the Born-Oppenheimer approach suggest that the lowest P wave excitations are below the threshold for Zweig allowed decay and thus should be extremely narrow.

\section{Summary}
\begin{itemize}
\item{The nature of the $B_s (1P) j^P=\frac{1}{2}^+$  states is still an open question. Comparing
the decays of the $D_s (1P)$ states and the $B_s(1P)$ states with give valuable
information into the microscopic nature of the $j^P=\frac{1}{2}^+$ 1P states.}
\item{The naive analog between the $Q\bar{q}$  mesons and the $QQq$ baryons fails for
the ccq, bbq, bcq systems. This is because the low-lying excitations of the QQ core are of the same order as that of the
excitations of the light quark ($\bar{q}$). Expect a complicated spectrum of low P wave QQq states.
Some of the lowest 1P (ccs, cbs, bbs) states may be stable to Zweig allowed decays.}
\end{itemize}
\section{Acknowledgements}
The collaboration of Chris Quigg on the work reported for doubly heavy baryons is acknowledged as well as the patience of  conference organizer Ayse Kizilersu.
This manuscript has been authored by Fermi Research Alliance, LLC under Contract No. DE-AC02-07CH11359
with the U.S. Department of Energy, Oﬃce of Science, Oﬃce of High Energy Physics. This research was supported by the 
Munich Institute for Astro and Particle Physics (MIAPP), which is funded by the Deutsche
Forschungsgemeinschaft (DFG, German Research Foundation) under Germany’s Excellence Strategy – EXC-2094 – 390783311. 
The author acknowledges the support of the Alexander von Humboldt Foundation  as well as
 the hospitality of the Munich Institute for Astro and Particle Physics (MIAPP) of the DFG cluster of excellence  ``Origin and Structure of the Universe" and also the Institute for Advanced Study at Technical University Munich where some of this research was performed.

 \bibliography{aus24}
 \bibliographystyle{unsrt}

\end{document}